\documentclass{optica-article}

\journal{opticajournal} 

\articletype{Research Article}
\usepackage{stfloats}
\usepackage{lineno}
\usepackage{pdfpages}

\begin{document}

\title{Photon-starved imaging through turbulence at the diffraction limit}

\author{Seungman Choi,\authormark{1} Peter Menart,\authormark{1} Andrew Schramka,\authormark{1} Shubhankar Jape,\authormark{1} Leif Bauer,\authormark{1} In-Yong Park,\authormark{2} and Zubin Jacob\authormark{1,*}}

\address{\authormark{1}Elmore Family School of Electrical and Computer Engineering, Birck Nanotechnology Center, Purdue University, West Lafayette, Indiana 47907, USA.
\\
\authormark{2}Emerging Research Instruments Group, Strategic Technology Research Institute, Korea Research Institute of Standards and Science (KRISS), 267 Gajeong-ro, Yuseong-gu, Daejeon 34113, South Korea
}

\email{\authormark{*}zjacob@purdue.edu} 


\begin{abstract*}
Atmospheric turbulence limits the resolution of ground-based imaging through rapidly evolving wavefront distortions. Fast acquisition can effectively freeze turbulence, but the restricted photon budget makes joint scene and wavefront recovery challenging. Here, we introduce Turbulence-Aware Poisson Blind Deconvolution (TAP-BD), which uses optical phase diversity to provide complementary measurements within a short acquisition window and increase wavefront information per detected photon. Poisson denoising and physics-informed optimization translate this information gain into joint recovery of the scene and a shared turbulence wavefront, with phase estimation errors approaching the Cramer--Rao lower bound in simulations. Experiments demonstrate reliable recovery from only a few tens of measurements under strong turbulence, with additional validation under severe photon scarcity. Simulations further show robustness to scintillation and temporal evolution over short acquisition windows. These results support photon-efficient imaging through turbulence, with potential applications in astronomy, remote sensing, and space situational awareness, particularly for time-sensitive observations.

\end{abstract*}

\section{Introduction}
High-resolution ground-based imaging in the visible and near-infrared is essential for space debris monitoring \cite{shell2010optimizing, menart2026definition, choi2024telescope, choi2026adaptive}, exoplanet detection \cite{sauvage2010analytical, bao_quantum-accelerated_2021}, and remote sensing \cite{lagouarde_atmospheric_2015}. Atmospheric turbulence limits spatial resolution by distorting the incoming wavefront. For targets in near-Earth orbit, target motion and tumbling also restrict acquisition time: measurements must be collected before changes in target position or appearance blur the recovered structure.

Existing approaches address atmospheric distortion through hardware correction or computational reconstruction. Adaptive optics (AO) \cite{norris_all-photonic_2020, go_meta_2024} and wavefront shaping \cite{candes_phase_2015, wu_wish_2019, lai_photoacoustically_2015, horstmeyer_guidestar-assisted_2015, mosk_controlling_2012} use wavefront sensing or calibration measurements to guide correction with deformable mirrors or spatial light modulators (SLMs). These approaches require strong sensing or calibration signals, which may be unavailable for faint targets without a suitable guide source. Computational methods recover the object directly from turbulence-distorted intensity measurements. Speckle imaging combines statistical correlations across short exposures \cite{katz_non-invasive_2014, chen_enhancing_2022, sun2024overcoming}, while multiframe blind deconvolution (MFBD) jointly estimates the object and the unknown point spread functions (PSFs) \cite{kohli2024wavefront, huebner_blind_2008, zhu_removing_2012, guo_richardsonlucy_2024}. Both can use changing atmospheric states to obtain measurement diversity. Collecting more measurements can improve reconstruction, but also extends the observation window, particularly for statistical speckle methods. The need for many frames can limit their applicability to dynamic scenes.

A shorter acquisition window limits both target changes and atmospheric evolution. At visible wavelengths, atmospheric coherence times are typically a few milliseconds \cite{griffiths2024comparison}. Fast optical modulation and detection at kHz rates \cite{deng_diffraction_2022,chao_high-brightness_2023,choi_time-multiplexed_2022} can collect multiple measurements before the wavefront changes substantially. Known phase-diversity patterns supply complementary information within this window, rather than waiting for turbulence to generate measurement diversity. The reconstruction then estimates one object and one shared turbulence wavefront from all measurements. This reduces the number of atmospheric unknowns while retaining the information needed for fast imaging. The trade-off is a lower total photon budget, which requires reconstruction to remain robust to photon noise.

For faint targets, photon noise can hinder reliable joint recovery of the object and the turbulence wavefront. Photon-limited phase retrieval and wavefront sensing have been extensively studied \cite{goy2018low, deng2020learning, kang2020phase, choi2026poisson}. In blind deconvolution, photon noise can obscure the measurement differences needed to distinguish object structure from wavefront distortion, destabilizing joint estimation \cite{sanghvi2022photon, sanghvi2023structured}. Reliable imaging within short acquisition windows therefore requires optical measurements that make efficient use of the information carried by the available photons, together with reconstruction that explicitly accounts for photon noise.

To address these acquisition and photon constraints, we introduce Turbulence-Aware Poisson Blind Deconvolution (TAP-BD), a hardware--software co-design framework for imaging through turbulence within short acquisition windows. TAP-BD uses known phase diversity in place of changing turbulence realizations to obtain complementary measurements while estimating a single shared wavefront. This reduces the number of atmospheric unknowns and increases the wavefront information available per detected photon. Poisson denoising and physics-informed optimization translate this information gain into reliable joint recovery of the scene and turbulence wavefront from few photon-limited measurements.

We demonstrate this capability through simulations and controlled experiments. Under fixed turbulence, TAP-BD recovers the scene and wavefront from only a few tens of measurements, with simulated phase errors approaching the Cramer--Rao lower bound. Simulations incorporating scintillation and temporal evolution evaluate the limits of the shared-wavefront approximation and show an advantage over the MFBD and bispectrum methods for short acquisitions. Photon-limited experiments further demonstrate the contribution of Poisson denoising to scene and wavefront recovery as the photon budget decreases.

\section{Optical Encoding and Joint Reconstruction}
\begin{figure*} [t]
\centering\includegraphics[width=1.0\textwidth]{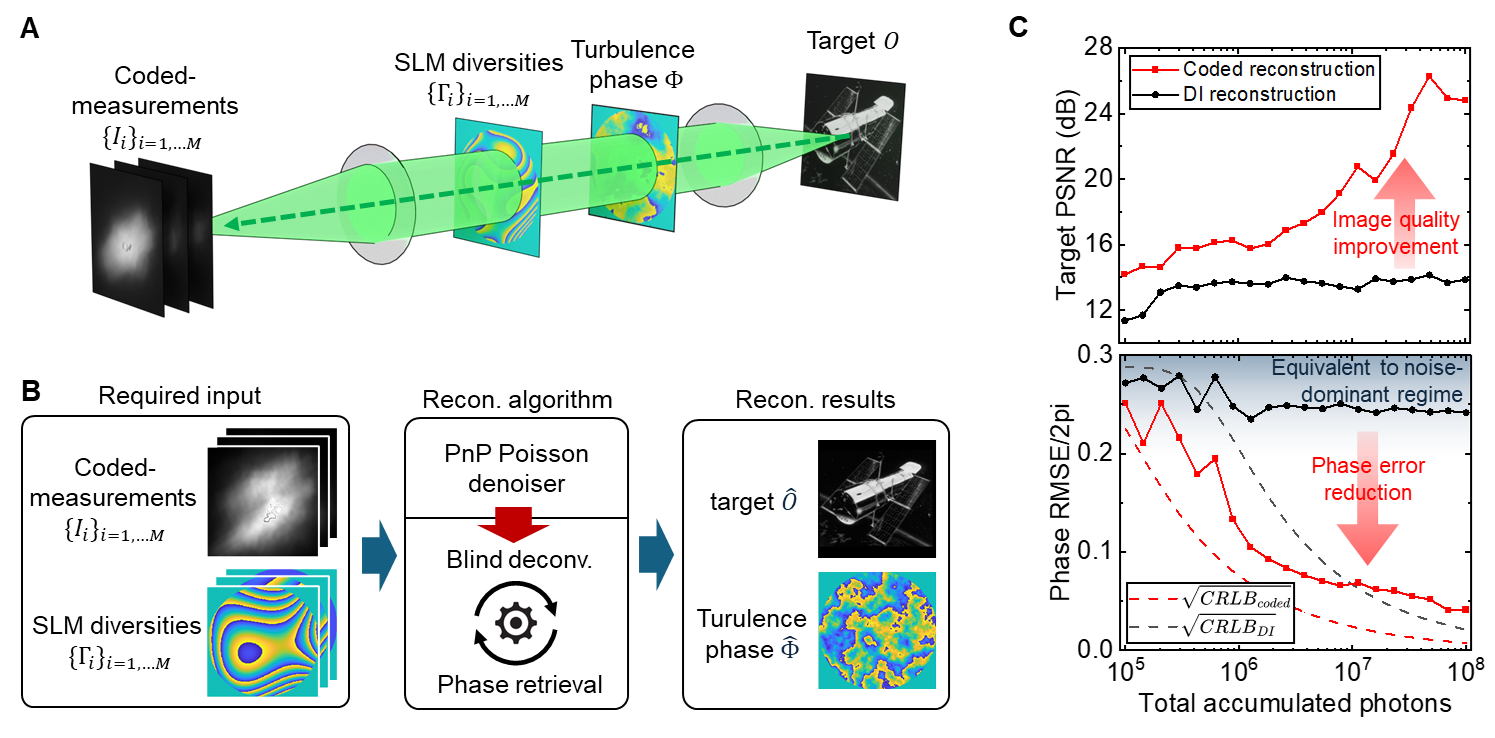}
\caption{
Turbulence-Aware Poisson Blind Deconvolution (TAP-BD) framework.
(A) Hardware setup: A spatial light modulator (SLM) sequentially applies known phase-diversity patterns ${\Gamma_i}$ to the distorted wavefront, generating coded measurements ${I_i}$.
(B) Reconstruction pipeline: A Poisson denoiser produces denoised measurements ${P_i}$, which an iterative solver combines with the known SLM patterns ${\Gamma_i}$ to jointly estimate the target $O$ and turbulence phase $\Phi$.
(C) Reconstruction performance using $M=10$ measurements. Target PSNR (top) and turbulence-phase RMSE (bottom) are compared for TAP-BD coded imaging (red) and direct imaging (DI; black). The dashed curves show the phase Cramer-Rao lower bounds (CRLBs), with the lower TAP-BD bound indicating more information content from phase-diverse encoding.
}
\label{fig1}
\end{figure*}

TAP-BD is designed to jointly recover the target scene and turbulence wavefront from a limited number of coded measurements within a short acquisition window. As illustrated in Fig.~\ref{fig1}A, atmospheric turbulence is approximated as stationary during short acquisition, allowing all measurements to constrain a single unknown turbulence phase $\Phi$.

An SLM sequentially applies $M$ known phase patterns ${\Gamma}_{i=1}^{M}$ to the turbulence-distorted light from the target $O$, generating distinct PSFs and corresponding intensity measurements. The applied patterns introduce measurement diversity without requiring atmospheric evolution, helping distinguish target structure from wavefront distortion. This encoding increases the Fisher information available for wavefront estimation under limited photon and measurement budgets, as quantified in Section~\ref{section3}.

The phase-diversity patterns are generated by combining the first 15 Zernike modes with Gaussian random coefficients of zero mean and variance 20.

Fig.~\ref{fig1}B illustrates the TAP-BD reconstruction pipeline. Each coded intensity measurement $I_i$ is first processed into a denoised measurement $P_i$ using an Anscombe variance-stabilizing transform and wavelet-domain total variation regularization \cite{shen_improved_2018}. This preprocessing suppresses shot-noise fluctuations that could otherwise produce spurious high-frequency structure in the object and wavefront estimates. By accounting for Poisson noise statistics, the denoiser adapts noise suppression to the photon level, limiting noise-driven spatial detail before joint reconstruction. The resulting denoised measurements $P_i$ serve as inputs to the joint reconstruction.

We model the atmospheric wavefront as a single pupil-plane phase screen that approximates the cumulative distortion from multiple atmospheric layers and is shared across the coded measurements. This representation neglects amplitude fluctuations under weak scintillation \cite{peters2025structured} and assumes negligible phase evolution over the short total acquisition window. Supplementary simulations support the phase-only approximation beyond the weak-scintillation regime, while the validity of the shared-wavefront approximation under temporal evolution is evaluated in Section~\ref{Sec_TimeVaryingSim}. Under monochromatic and isoplanatic imaging conditions, the joint estimation problem is formulated as:

\begin{equation}
L(O,\Phi) = \sum_{i=1}^{M} \frac{1}{2} \left\lVert O * \left| \mathcal{F}_i \Phi \right|^2 - P_i \right\rVert^2
\label{Eq1}
\end{equation}

where $*$ denotes convolution, $P_i$ is the Poisson-denoised intensity measurement corresponding to the $i$-th coded frame, and $\mathcal{F}_i$ is the propagation operator incorporating the known phase-diversity pattern $\Gamma_i$, implemented using the angular spectrum method (see Appendix).
Direct optimization of Eq.~\ref{Eq1} can be challenging because the object and turbulence are coupled through nonlinear image formation, particularly under strong turbulence or limited measurements. 

To stabilize joint reconstruction, we introduce the PSFs ${h_i}$ as auxiliary variables and use ADMM to alternate between blind deconvolution and phase retrieval \cite{antipa2017diffusercam}. The blind deconvolution step estimates the object $O$ and PSFs from the denoised measurements, with regularization to suppress residual noise. The phase retrieval step then estimates a single turbulence wavefront that explains these PSFs under the known diversity patterns. Applying regularization and wavefront constraints at successive steps helps limit the propagation of noise between the object and wavefront estimates.

Specifically, the object $O$ and intermediate PSFs ${h_i}$ are updated using the blind deconvolution objective:

\begin{equation}
L_1(O,{h_i},T_O) =
\sum_{i=1}^{M}\frac{1}{2}\left\lVert O*h_i-P_i\right\rVert^2
+\lambda\left\lVert T_O\right\rVert_1,
\quad T_O=\Psi O
\label{eq2}
\end{equation}

where $\Psi$ is the total variation (TV) operator and $\lambda$ controls object regularization. Using an augmented Lagrangian formulation, $O$, ${h_i}$, and $T_O$ are updated alternately, with convolution evaluated as pointwise multiplication in the Fourier domain.

The shared turbulence phase $\Phi$ is then updated using the phase retrieval objective:

\begin{equation}
L_2({u_i},\Phi)=
\sum_{i=1}^{M}\frac{1}{2}
\left\lVert\sqrt{h_i}-|\mathcal{F}_i\Phi|\right\rVert^2,
\quad u_i=\mathcal{F}_i\Phi
\end{equation}

where $u_i$ is the detector-plane field associated with the $i$-th diversity pattern. This objective matches the predicted PSF amplitudes to those obtained from blind deconvolution, while requiring all PSFs to be explained by a single shared wavefront under the known diversity patterns.

The PSFs are subsequently recomputed as  $h_i=|\mathcal{F}_i\Phi|^2$ for the next blind deconvolution update. Full update rules are provided in the Supplementary Information.

\section{Information-Theoretic Gain with TAP-BD
\label{section3}}
Under a limited photon budget, optical encoding can improve reconstruction by increasing the information obtained from the available photons. Fig.~\ref{fig1}C quantifies this information gain by comparing TAP-BD with direct imaging (DI), an uncoded baseline under the same photon budget. The dashed curves show the Cramer--Rao lower bounds (CRLBs) for turbulence phase estimation, derived from the Fisher information to characterize the minimum variance attainable by unbiased estimators. TAP-BD consistently yields a lower CRLB, indicating greater information about the unknown wavefront even when the object is also unknown, without increasing the photon budget. This is consistent with prior findings that wavefront randomization can improve parameter distinguishability \cite{kohli2024wavefront}.

The reconstruction results (solid curves) show that the information gain translates into improved estimation accuracy. TAP-BD achieves turbulence phase errors approaching the corresponding CRLB across the tested photon levels, together with higher target peak signal-to-noise ratio (PSNR) than DI. Details of the performance metrics and CRLB calculations are provided in the Appendix.

\begin{figure*}[t]
\centering\includegraphics[width=1.0\textwidth]{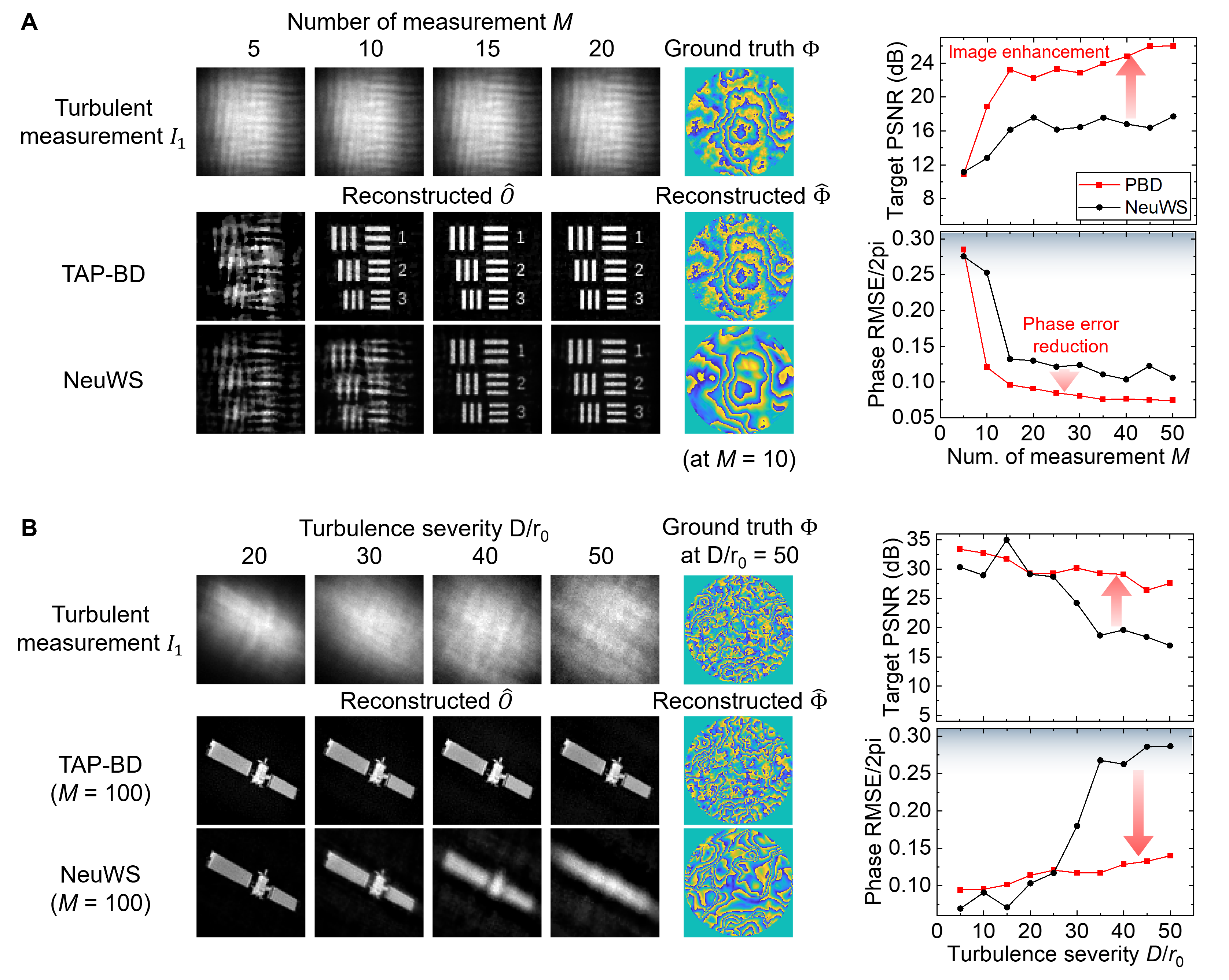}
\caption{
Comparative simulations of TAP-BD and NeuWS. (A) Performance versus measurement count $M$ at fixed turbulence strength $D/r_0=20$. (B) Performance versus turbulence strength $D/r_0$ at fixed $M=100$. The left panels show a turbulence-distorted measurement $I_1$ acquired with a flat SLM pattern $\Gamma_1$, along with the reconstructed targets $\hat{O}$ and estimated turbulence phases $\hat{\Phi}$. The right panels show the corresponding target PSNR and phase RMSE. For a consistent comparison, NeuWS was evaluated using the same pupil model as TAP-BD, with uniform amplitude over a circular aperture and phase-only aberrations. NeuWS employed a five-layer neural network with a hidden width of 32. Both methods were run for 2000 iterations (epochs for NeuWS), with a photon budget of $1\times10^7$ photons per frame. Poisson denoising was disabled for TAP-BD in these simulations.
}
\label{fig2}
\end{figure*}

\section{Measurement-Efficient Reconstruction under Strong Turbulence}

Fig.~\ref{fig2} evaluates TAP-BD as a function of measurement count $M$ and turbulence strength $D/r_0$, where $D$ is the aperture diameter and $r_0$ is the Fried parameter (see Appendix). We compare TAP-BD with NeuWS \cite{feng_neuws_2023}, which represents aberrations through a neural network with Zernike feature encoding. TAP-BD instead estimates the turbulence phase directly on the pupil grid using ADMM-based alternating optimization. Both methods use identical coded measurements and photon budgets, with turbulence held static within each sequence. The effect of atmospheric evolution and comparisons with conventional astronomical imaging methods are examined in Section~\ref{Sec_TimeVaryingSim}.

At fixed $D/r_0=20$, TAP-BD achieves accurate reconstruction with as few as $M=10$ measurements, with target PSNR increasing from approximately 18 to 26 dB as $M$ increases (Fig.~\ref{fig2}A). NeuWS also recovers recognizable target structures, but its reconstructions remain blurred, with PSNR saturating near 17 dB. In wavefront estimation, TAP-BD better preserves high-frequency features, achieving a phase RMSE$/2\pi$ below 0.1, compared with approximately 0.12 for NeuWS.

At fixed $M=100$, TAP-BD maintains high reconstruction accuracy across turbulence strengths up to $D/r_0=50$ (Fig.~\ref{fig2}B). NeuWS, in contrast, progressively loses fine structural details beyond $D/r_0\approx30$, accompanied by decreasing target PSNR and increasing phase RMSE. Together, these results demonstrate that TAP-BD can reduce the number of measurements and total photons required for accurate recovery while achieving higher reconstruction fidelity under the same measurement and photon budgets.

TAP-BD is also computationally efficient, using FFT-based updates with a per-iteration complexity of $O(MN_{\mathrm{pix}}\log N_{\mathrm{pix}})$, where $N_{\mathrm{pix}}$ is the number of pixels in the reconstruction grid. For $M=20$ and a $256\times256$ grid, it requires 18.4 s on an RTX 3060Ti GPU, compared with 152 s for NeuWS (Table~\ref{tab_excutetime}).

\begin{table}[t]
\centering
\caption{Execution time comparison between TAP-BD and NeuWS (each using 2000 iterations/epochs) on a standard GeForce RTX 3060Ti GPU.}
\label{tab_excutetime}
\begin{tabular}{c c c}
\hline
\textbf{Number of measurements ($M$)} & \textbf{TAP-BD (s)} & \textbf{NeuWS (s)} \\
\hline
10   & 12.0  & 98  \\
20   & 18.4  & 152  \\
40   & 32.4 & 274 \\
100  & 77.2 & 688 \\
\hline
\end{tabular}
\end{table}

\section{\label{Sec_TimeVaryingSim}Reconstruction under Time-Varying Atmospheric Turbulence}

TAP-BD models the turbulence phase as shared across the coded measurements acquired within a short sequence. To evaluate this approximation under atmospheric evolution, we simulated three frozen-flow turbulence layers with a combined nominal turbulence strength of $D/r_0=16$, with propagation between the layers producing both phase distortions and scintillation in the pupil plane. Measurements were simulated assuming a 1 kHz SLM update rate and a 1 ms exposure per frame, accounting for turbulence evolution both between frames and within each exposure. Each frame contained $5\times10^6$ photons. Further simulation details are provided in the Supplementary Information.

Figure~\ref{fig2_add}A shows how atmospheric coherence time $\tau_0$ determines the useful acquisition window for TAP-BD, comparing sequences of $M=10$, $20$, and $50$ measurements. At the longest coherence time tested ($\tau_0=50$ ms), the longest sequence ($M=50$) provides the highest object PSNR by accumulating more phase-diverse measurements and photons. As $\tau_0$ decreases, however, atmospheric evolution increasingly violates the shared-wavefront approximation, causing PSNR to decline sharply for longer sequences. Shorter sequences ($M=10$) limit this mismatch and become favorable under faster turbulence. The preferred sequence length therefore depends on the balance between measurement accumulation and wavefront evolution.

Figure~\ref{fig2_add}B compares TAP-BD with MFBD using coded or uncoded measurements \cite{asensio2025torchmfbd} and bispectrum speckle imaging \cite{goodman199310,herring2016fast} at fixed $\tau_0=10$ ms. TAP-BD reaches a peak object PSNR above $26$ dB within approximately $10$--$20$ ms, showing that accurate object reconstruction remains possible even when the acquisition window moderately exceeds $\tau_0$. Although its PSNR subsequently decreases as more measurements are acquired over a longer acquisition window, TAP-BD retains an advantage over the evaluated benchmarks for short acquisitions, with the best-performing benchmark reaching comparable PSNR near 80 ms under these conditions. These results indicate that TAP-BD is well suited to imaging with limited acquisition time. Representative reconstructions are provided in the Supplementary Information.

\begin{figure*}
\centering\includegraphics[width=0.7\textwidth]{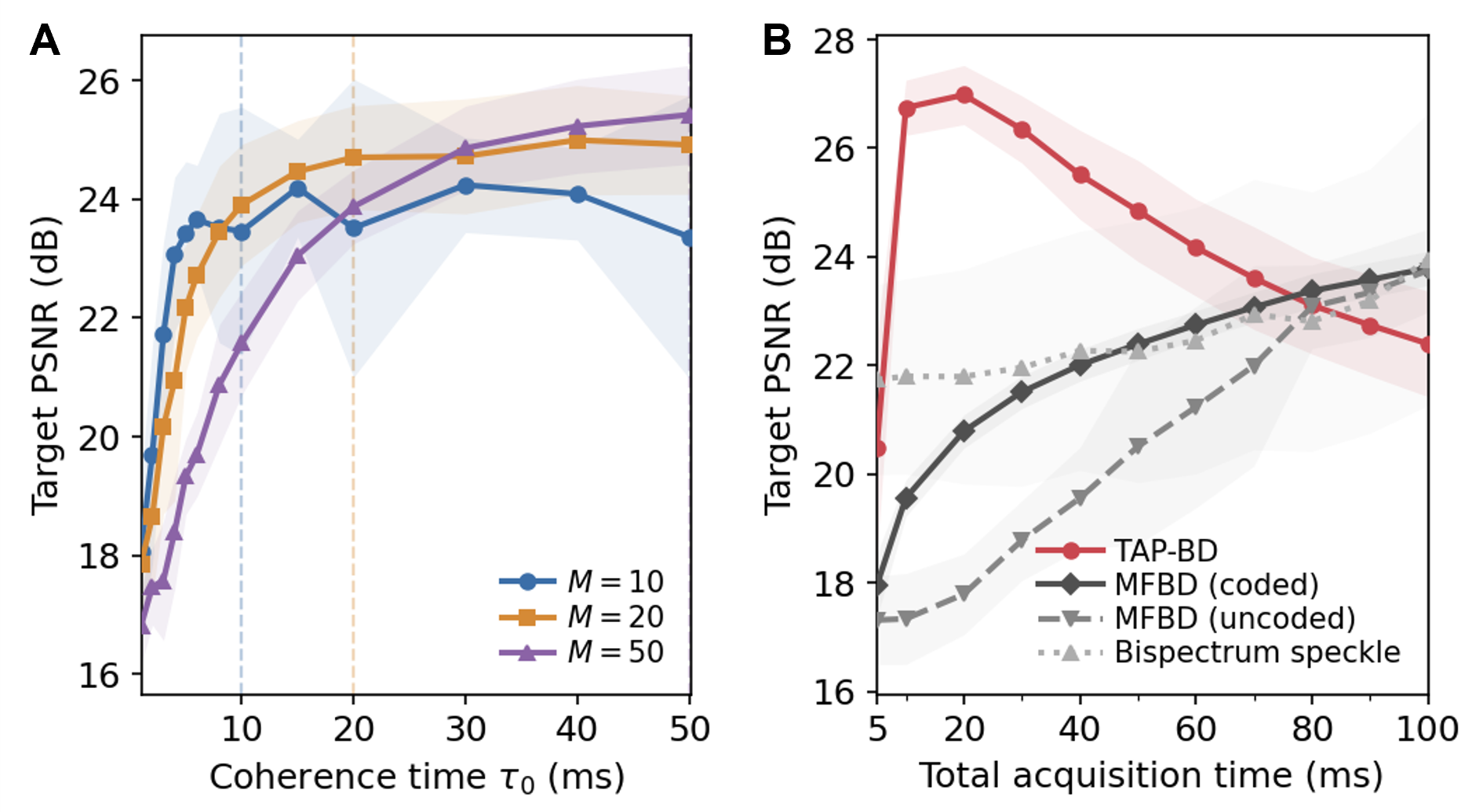}
\caption{
Reconstruction performance under time-varying atmospheric turbulence at nominal $D/r_0=16$. Measurements are simulated assuming a 1 kHz SLM update rate and a 1 ms exposure per frame.
(A) Target object PSNR versus atmospheric coherence time $\tau_0$ for $M=10$, $20$, and $50$ coded measurements. Vertical dashed lines indicate $\tau_0=M\times1$ ms, where the acquisition duration equals the coherence time.
(B) Target object PSNR versus total acquisition time at fixed $\tau_0=10$ ms, comparing TAP-BD with MFBD using coded or uncoded measurements and bispectrum speckle imaging. Coded measurements include known phase-diversity patterns.
Simulated measurements incorporate evolving phase and amplitude fluctuations, with $5\times10^6$ photons per frame, while TAP-BD estimates a single shared turbulence phase. Curves and shaded regions show the mean and $\pm$ one standard deviation over 20 independent turbulence realizations. Simulation details and representative reconstructions are provided in the Supplementary Information.
}
\label{fig2_add}
\end{figure*}

\begin{figure*}
\centering\includegraphics[width=1.0\textwidth]{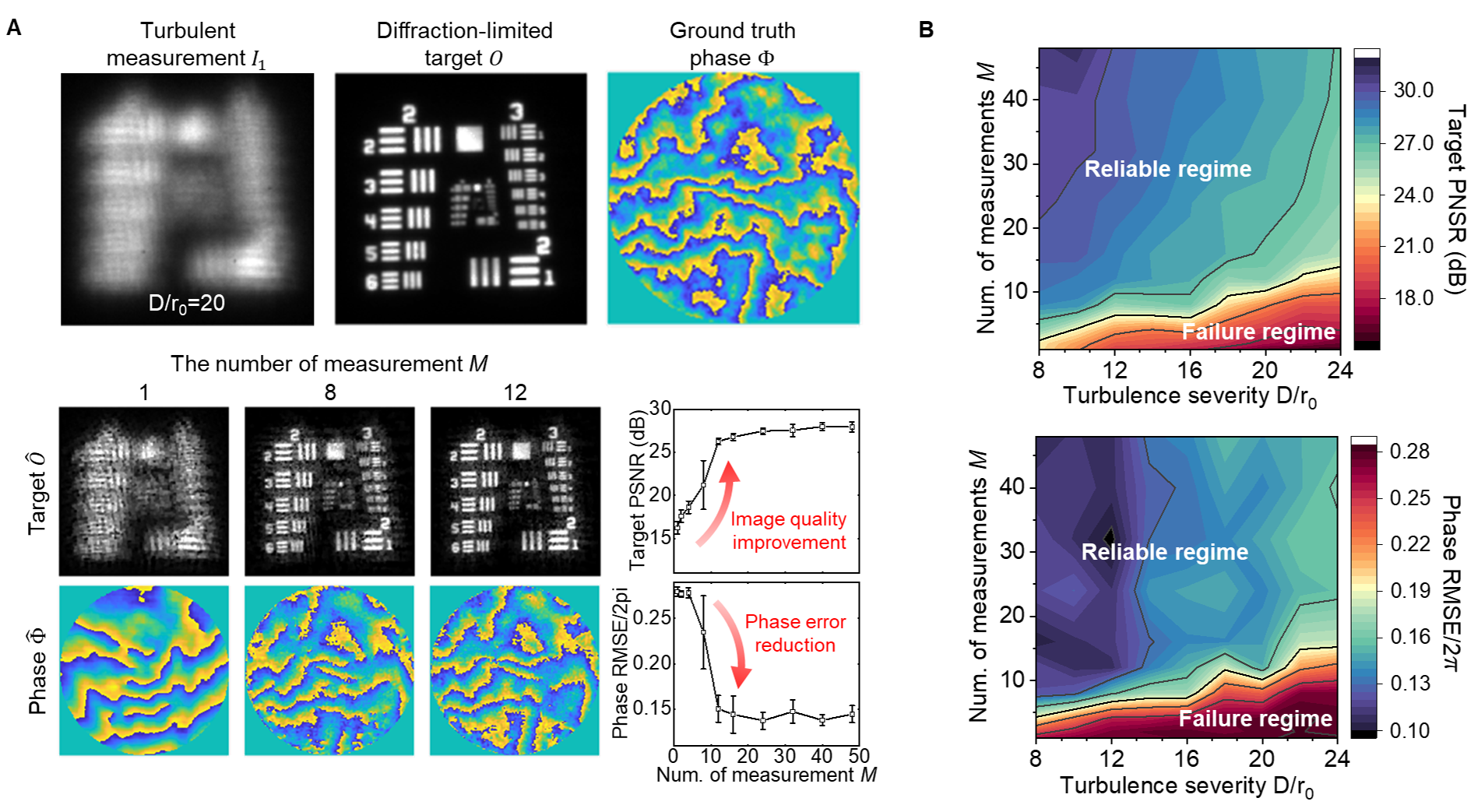}
\caption{
Experimental reconstruction under varying turbulence severity level $D/r_0$ and measurement count M, averaged over five distinct turbulence phases per $D/r_0$. (A) Example reconstruction results at $D/r_0=20$: The top panel shows the turbulent measurement $I_1$ with flat phase diversity $\Gamma_1$, diffraction-limited image of the USAF 1951 target O and ground truth turbulence phase $\Phi$. The bottom plot illustrates how increasing $M$ improves reconstruction quality, as measured by target PSNR, and phase RMSE (bottom-right). (B) Contour plots of the target PSNR (top) and phase RMSE (bottom) as functions of $D/r_0$ and $M$. All experiments were conducted with a high photon flux ($\sim2.13\times10^7$ photons per single measurement), using a 5 mm aperture and 1000 iterations per reconstruction. A Poisson denoiser was not employed.
}
\label{fig3}
\end{figure*}

\section{Experimental Validation under Controlled Turbulence}

We experimentally evaluate TAP-BD across varying turbulence strengths $D/r_0$ and measurement counts $M$ (Fig.~\ref{fig3}). A negative 1951 USAF resolution target (R3L3S1N, Thorlabs) serves as the test object $O$. An SLM (E19×12-400-800-HDM8, Meadowlark Optics) generates a fixed turbulence phase $\Phi$ for each measurement sequence, superposed with the known phase diversity patterns ${\Gamma_i}$. The resulting coded measurements ${I_i}$ are recorded using an emCCD camera (PIMAX 4:512EM, Teledyne Princeton Instruments). Further experimental details and photon calibration are provided in the Supplementary Information and Appendix, respectively.

Fig.~\ref{fig3}A shows representative reconstruction results at $D/r_0=20$. With only $M=12$ measurements, TAP-BD achieves approximately 26 dB target PSNR and a turbulence phase RMSE$/2\pi$ of 0.15. The recovered target resolves features approaching the diffraction limit of the experimental system, corresponding to Group 3, Element 4 of the USAF target (11.31 line pairs/mm, approximately 44 $\mu$m).

Fig.~\ref{fig3}B quantifies the relationship between turbulence strength and the number of measurements required for reliable reconstruction. As $D/r_0$ increases, more measurements are generally needed to maintain reconstruction fidelity. Nevertheless, TAP-BD achieves near-diffraction-limited reconstruction with fewer than 15 measurements for $D/r_0\leq24$, resolving Group 3, Element 4 of the USAF target, at a target PSNR of approximately 24 dB. These results demonstrate that TAP-BD can recover both the scene and wavefront from a short measurement sequence even under strong turbulence.

\begin{figure*}
\centering\includegraphics[width=1.0\textwidth]{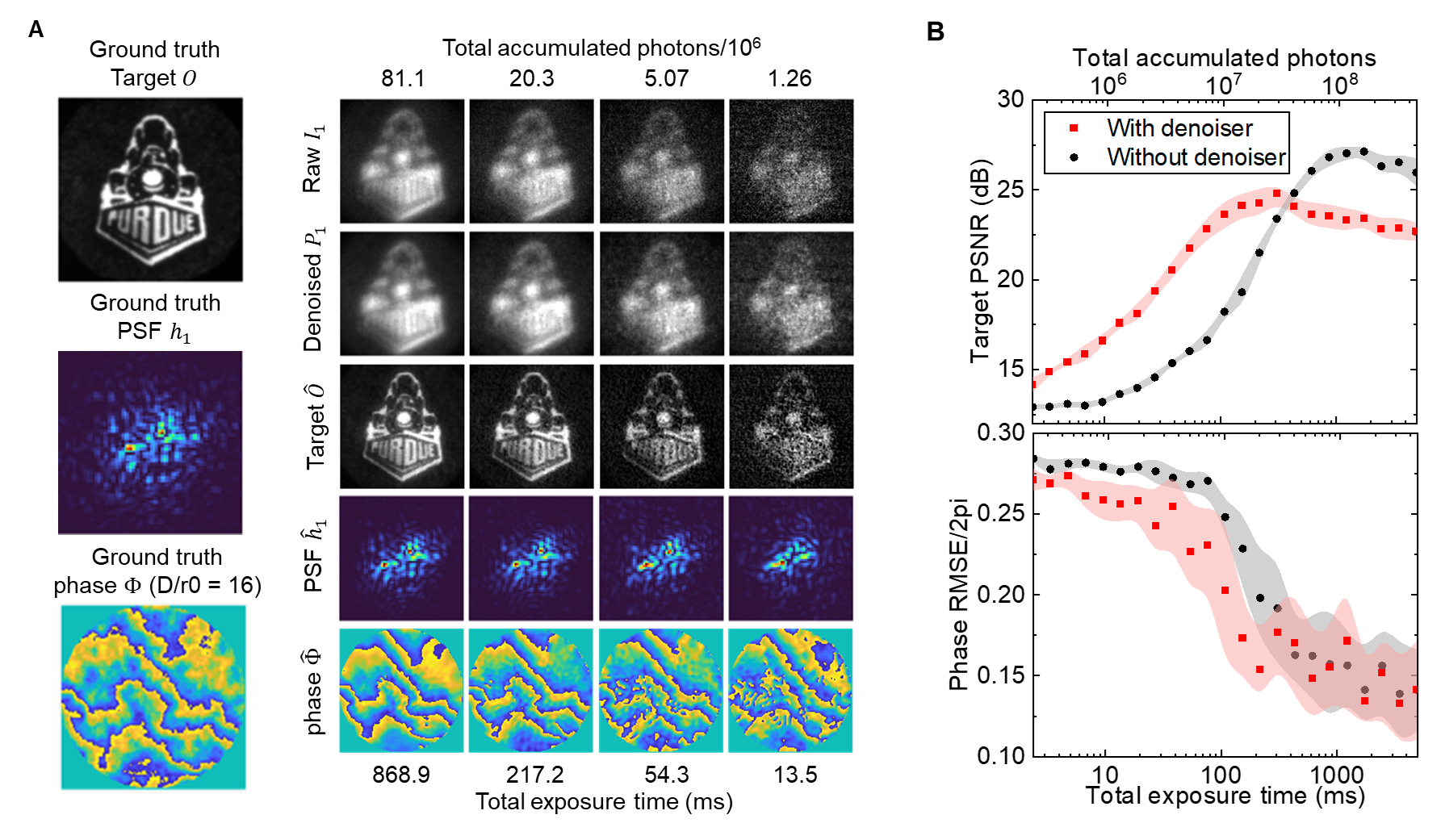}
\caption{
Experimental reconstruction under low photon levels, averaged over five distinct turbulence phases at fixed $D/r_0=16$. (A) Example reconstructions. The left panel shows ground truth elements: the diffraction limited image of ‘Purdue train’ target $O$, the turbulent PSF $h_1$ with flat phase diversity $\Gamma_1$ and the turbulence phase $\Phi$. The right panels illustrate reconstruction results at different photon levels. The first row shows raw measurements $I_1$, highlighting how shot noise becomes dominant as photon count decreases. The second row presents Poisson-denoised measurements $P_1$. Rows three to five display the reconstructed target $\hat{O}$, estimated PSF $\hat{h_1}$ and retrieved phase $\hat{\Phi}$. (B) Target PSNR and phase RMSE with and without Poisson denoising. In the low-photon regime, denoising significantly boosts PSNR and lowers phase RMSE. Shaded regions denote the $\pm$ one standard deviation computed over five independent trials with different turbulence phases. 
All experiments used a 3 mm aperture, $M=48$ measurements, and 1000 iterations per reconstruction. Note that exposure time here was adjusted solely to control photon levels, not to reflect a physical turbulence time scale.
}
\label{fig4}
\end{figure*}

\section{Robustness of TAP-BD to Photon-Starved Conditions}

We next evaluate TAP-BD under photon-starved conditions, where Poisson noise poses an additional challenge to joint scene and wavefront reconstruction (Fig.~\ref{fig4}). Experiments were conducted using a laser-printed ``Purdue train'' target at $D/r_0=16$, with $M=48$ coded frames. The total photon budget across these frames was varied from approximately $1.26\times10^6$ to $8.11\times10^7$ photons, corresponding to an average of 0.54 to 34.9 photons per pixel per frame over the $220\times220$-pixel EMCCD region. Photon calibration details are provided in the Appendix.

Fig.~\ref{fig4}A shows representative measurements and reconstructions across different photon levels. As the photon budget decreases, shot noise becomes increasingly pronounced in the raw measurements $\{I_i\}$, reducing the information available to distinguish scene structure from turbulence distortion. Despite this loss of measurement quality, TAP-BD continues to recover the target $\hat{O}$, PSF $\hat{h}_1$, and turbulence phase $\hat{\Phi}$ across the tested photon range. Fine spatial details gradually diminish at lower photon levels, but the dominant features of both the scene and wavefront remain identifiable. The gradual degradation in reconstruction fidelity is consistent with the photon-dependent estimation limits illustrated in Fig.~\ref{fig1}C.

Fig.~\ref{fig4}B further examines the contribution of Poisson denoising to joint reconstruction. At low photon counts, denoising improves target PSNR and reduces phase RMSE by mitigating noise in the coded measurements before joint optimization. As the photon budget increases, the difference between reconstructions with and without denoising becomes smaller, indicating that its contribution is most significant when shot noise dominates. Supplementary experiments demonstrate reconstruction of additional target geometries and optical wavefront correction using the recovered phase. At low photon levels, Poisson denoising improves not only computational reconstruction but also the optical correction obtained from the estimated wavefront.

\section{Discussion}

TAP-BD shows that optical measurement diversity can reduce the need to sample many atmospheric states for joint scene and wavefront recovery. By introducing known phase patterns within a short acquisition window, the method obtains complementary constraints while estimating a single shared turbulence wavefront. Its robustness to photon scarcity makes this strategy useful when extending acquisition to collect more light would conflict with atmospheric evolution or target dynamics.

TAP-BD estimates only the pupil phase, reducing the number of unknowns but leaving amplitude fluctuations caused by scintillation unmodeled.  Supplementary simulations show that this simplified model retains effective scene reconstruction and wavefront correction even beyond weak scintillation. Under stronger scintillation, a compact, physically constrained representation of pupil-amplitude fluctuations could extend its applicability \cite{feng_neuws_2023, martin2025single}.

The shared-wavefront approximation reduces the number of atmospheric unknowns but introduces a trade-off between measurement accumulation and model mismatch. Figure~\ref{fig2_add} shows that additional measurements improve reconstruction when atmospheric evolution is slow, whereas shorter sequences become preferable as the wavefront changes more rapidly. Acquisition duration should therefore be chosen to retain sufficient measurement diversity while limiting wavefront evolution. Allowing the wavefront to evolve across the acquisition window could accommodate longer acquisitions, with a compact temporal representation limiting the number of additional atmospheric variables \cite{jefferies2011deconvolution,ren2022constrained}.

The current model assumes an isoplanatic response and monochromatic illumination. Extension to wider fields of view could incorporate spatially varying or multiplanar propagation models to account for field-dependent aberrations \cite{pellizzari_imaging_2019, yanny_deep_2022, shajkofci_spatially-variant_2020}, while broader spectral bands could be addressed through wavelength-dependent image formation that accounts for chromatic PSF variations \cite{huang_spectral_2022}.

Overall, TAP-BD combines informative optical encoding with photon-noise-aware reconstruction to recover target structure and atmospheric aberrations from few measurements. This enables imaging within short acquisition windows despite limited photon budgets, with potential applications in time-resolved astronomy and rapid space-object imaging for space situational awareness.

\bigskip

\section*{Appendix A: Target PSNR and phase RMSE calculation}
Joint object and wavefront estimation involves an inherent ambiguity between tip-tilt and target position. For PSNR and phase RMSE evaluation, we resolve this ambiguity using the known ground-truth target position. 
Specifically, for the target PSNR calculation, we first determine the lateral shift between the reconstructed target $\hat{O}$ and the ground truth $O$ by computing their cross-correlation:
\begin{equation}
C(\Delta x,\Delta y) = \sum_{x,y} O(x,y)\,\hat{O}(x+\Delta x, y+\Delta y)
\tag{A1}
\end{equation}
where $(x,y)$ is detector pixel coordinate. 

The optimal shift $(\Delta x^*, \Delta y^*)$ corresponding to the peak of the correlation is then given by:

\begin{equation}
(\Delta x^*, \Delta y^*) = \text{argmax} [C(\Delta x,\Delta y)]
\tag{A2}
\end{equation}

Using this shift, we align the reconstructed target:
$\hat{O}^*(x,y) = \hat{O}(x+\Delta x^*, y+\Delta y^*)$,
which allows us to compute the target PSNR:
\begin{equation}
\mathrm{PSNR} = 10\log_{10}\!\left(\frac{\max(O)^2}{\tfrac{1}{K}\sum_{x,y}[O(x,y)-\hat{O}^*(x,y)]^2}\right),
\tag{A3}
\end{equation}
where $K$ is the total number of target pixels.

The obtained lateral shift is converted into an equivalent tip/tilt phase correction. Assuming a linear phase ramp, the tip/tilt correction can be expressed as:
\begin{equation}
\Delta(x,y) = \frac{2\pi}{\lambda}\,\frac{s}{f}\left(\Delta x^*\,x+\Delta y^*\,y\right),
\tag{A4}
\end{equation}
where $\lambda$ is the wavelength, $s$ is detector pixel size, and $f$ is focal length of imaging lens. 
The reconstructed phase $\hat{\Phi}$ is then corrected by adding the tip/tilt components:
$\hat{\Phi}^*(x,y) = \hat{\Phi}(x,y)\exp\!\left(j\,\Delta(x,y)\right)$.

Because phase is inherently cyclic, the standard RMSE does not properly account for phase wrapping. Before evaluating the error, we align the global phase offset (piston) to minimize the wrapped phase RMSE. Here, $\hat{\Phi}^*$ denotes the estimate after both tip/tilt and piston alignment. The absolute phase error between $\angle\Phi$ and $\angle\hat{\Phi}^*$ is defined as:
\begin{equation}
E(\Phi,\hat{\Phi}^*) =
\begin{cases}
|\angle\Phi - \angle\hat{\Phi}^*|, & \text{if } |\angle\Phi - \angle\hat{\Phi}^*| \leq \pi, \\
2\pi - |\angle\Phi - \angle\hat{\Phi}^*|, & \text{otherwise}.
\end{cases}
\tag{A5}
\end{equation}

The piston-aligned phase RMSE is then calculated as:
\begin{equation}
\mathrm{RMSE} = \sqrt{\frac{1}{K}\sum_{x,y}\big[E(\Phi,\hat{\Phi}^*)\big]^2},
\tag{A6}
\end{equation}
where $K$ is the total number of phase pixels within the pupil. Piston alignment makes this metric invariant to a global phase offset.

\section*{Appendix B: Kolmogorov turbulence theory}

Kolmogorov turbulence theory provides a realistic approach for simulating atmospheric turbulence, particularly when rapid turbulence variations make the Zernike-based approximation unsuitable. To generate a single turbulence phase map $\Phi(\mathbf{x})$, where $\mathbf{x}=(x,y)$, we define the turbulence severity $D/r_0$, with $D$ as the aperture diameter at the pupil-plane and $r_0$ as the Fried parameter representing the maximum distance for coherent light transmission through the turbulence:

\begin{equation}
r_0 = \left( 0.423 k^2 \int_{L} C_n^2(z)\, dz \right)^{-3/5}
\tag{B1}
\end{equation}
where $k$ is the optical wavenumber and $C_n^2$ is the refractive index structure constant over the propagation path $L$. 

The power spectral density (PSD) $P(\mathbf{k})$, describing phase fluctuations across spatial frequencies $\mathbf{k}=(k_x,k_y )$, follows Kolmogorov’s model:

\begin{equation}
P(\mathbf{k}) = 0.4961 \, r_0^{-5/3} \, \mathbf{k}^{-11/3}
\tag{B2}
\end{equation}

We define the complex turbulence spectrum with random phase distribution:

\begin{equation}
S_{\mathrm{turb}}(\mathbf{k}) = \sqrt{P(\mathbf{k})}\,\eta(\mathbf{k})
\tag{B3}
\end{equation}
where $\eta(\mathbf{k})$ is complex Gaussian noise with zero mean and standard deviation $\sigma = 1$. 

The turbulence phase map $\Phi(\mathbf{x})$ is obtained via the inverse Fourier transform:

\begin{equation}
\Phi(\mathbf{x}) = \mathbf{F}^{-1}\!\left[ S_{\mathrm{turb}}(\mathbf{k}) \right]
\tag{B4}
\end{equation}

This phase screen $\Phi(\mathbf{x})$ simulates the effect of random turbulence over $L$ for a specific $r_0$.

\section*{Appendix C: Wave propagator}
To simulate the e-field propagation over a distance $z$, we use an angular spectrum method (ASM), denoted as $A_z$.  
The ASM operator is defined as:
\begin{equation}
e_{\mathrm{out}}(x,y) = \mathcal{A}_z e_{\mathrm{in}}(x,y) = 
\mathbf{F}^{-1}\!\left[ \mathbf{H}(k_x,k_y,z) \cdot \mathbf{F}\big[e_{\mathrm{in}}(x,y)\big] \right]
\tag{C1}
\end{equation}
where $e_{\mathrm{in}}(x,y)$ is the input e-field, $e_{\mathrm{out}}(x,y)$ is the propagated field, and $\mathbf{F}$ and $\mathbf{F}^{-1}$ represent the Fourier and inverse Fourier transform, respectively.

The transfer function $\mathbf{H}(k_x,k_y,z)$ for distance $z$ is:
\begin{equation}
\mathbf{H}(k_x,k_y,z) = \exp\!\left( \frac{2 j \pi z}{\lambda} \sqrt{\,1 - \lambda^2 (k_x^2 + k_y^2)} \,\right)
\tag{C2}
\end{equation}
Here, $k_x$ and $k_y$ are spatial frequency components, and $\lambda$ is the wavelength.  
Above expression holds if $k_x^2 + k_y^2 \leq 1$, and is $0$ otherwise.

For simulating field propagation between sequential optical elements, such as from the SLM to an imaging lens, and from the lens to a detector, we use:
\begin{equation}
u_i = \mathcal{F}_i(\Phi) = \mathcal{A}_f \left\{ \mathcal{A}_f \left[ A \cdot \Phi \cdot \Gamma_i \right] \cdot R \right\}
\tag{C3}
\end{equation}
where $u_i$ is the resulting image plane e-field, $\Phi$ is the pupil plane turbulence e-field (the target field to retrieve), $A$ is the aperture constraint, $\Gamma_i$ is the $i$-th phase diversity pattern applied by the SLM, and $R$ is the imaging lens phase with focal length $f$.

\section*{Appendix D: CRLB calculation}
To quantify the theoretical phase estimation limit shown in Fig.~\ref{fig1}C, we computed the Cramer-Rao lower bound (CRLB) for joint estimation of the turbulence phase and target intensity. The target is treated as unknown, so the phase bound accounts for uncertainty in the target rather than assuming that it is known.

The full parameter vector is
$\Theta = \left(\Phi_1,\ldots,\Phi_A,O_1,\ldots,O_B\right)^{\mathsf T}$,
where $A=1184$ is the number of phase pixels within a circular aperture of 40 pixels in diameter, and $B=1600$ is the number of target intensity parameters in a $40\times40$ image. We consider $M=10$ coded measurements, each containing $K$ detector pixels.

For independent Poisson measurements, the Fisher information matrix (FIM) per unit photon budget is
\begin{equation}
J_{pq} =
\sum_{k=1}^{K}\sum_{i=1}^{M}
\frac{1}{\mu_{ik}}
\left(\frac{\partial\mu_{ik}}{\partial\Theta_p}\right)
\left(\frac{\partial\mu_{ik}}{\partial\Theta_q}\right),
\tag{D1}
\end{equation}
where $\mu_{ik}$ is the expected intensity per unit photon budget at detector pixel $k$ in measurement $i$. The indices $p$ and $q$ span both phase and target intensity parameters. Thus, the FIM includes the phase block $J_{\Phi\Phi}$, the target intensity block $J_{OO}$, and the cross-information blocks $J_{\Phi O}$ and $J_{O\Phi}$.

The measurements are invariant to a global phase shift, making the FIM singular due to the piston ambiguity. We fix the phase at one pupil pixel and remove its corresponding row and column from the full FIM \cite{bandeira_saving_2014}. All target intensity parameters remain unknown. We then invert the reduced joint FIM, denoted by $J_{\mathrm{red}}$, and extract its phase block:
\[
C_{\Phi} =
\left[J_{\mathrm{red}}^{-1}\right]_{\Phi\Phi}.
\]
The resulting phase bound accounts for uncertainty in the jointly estimated target intensity.

For photon budget $N$, the FIM scales as $N J_{\mathrm{red}}$, and the corresponding phase covariance bound is $C_{\Phi}/N$. We summarize this bound using the root-mean-square phase standard deviation:
\begin{equation}
\bar{\sigma}(N) =
\sqrt{\frac{1}{N(A-1)}
\sum_{a=1}^{A-1}(C_{\Phi})_{aa}},
\tag{D2}
\end{equation}
which gives the phase RMSE lower bound for locally unbiased joint estimation in the chosen phase reference.

\section*{Appendix E: Photon number measurement}
We calculate the photon count using two methods. 
The first method is based on the EM-CCD measurement, where the photon count $N_{\mathrm{EM}}$ is computed from the intensity image $I(x)$ as:
\begin{equation}
N_{\mathrm{EM}} = \sum_{x} \frac{I(x)}{Q_{\mathrm{EM}} \cdot G}
\tag{E1}
\end{equation}
where $Q_{\mathrm{EM}}$ is the quantum efficiency of the EM-CCD, and $G$ is the system gain.

The second method uses a photodiode (S120VC, Thorlabs) to obtain a reference photon counts $N_{\mathrm{PM}}$:
\begin{equation}
N_{\mathrm{PM}} = P \cdot\frac{\lambda}{h c} \cdot \frac{1}{Q_{\mathrm{PM}}}
\tag{E2}
\end{equation}
where $P$ is the measured power, $hc/\lambda$ is the energy per photon, and $Q_{\mathrm{PM}}$ is the quantum efficiency of the photodiode.

Upon comparing $N_{\mathrm{PM}}$ and $N_{\mathrm{EM}}$, we found that $N_{\mathrm{PM}}$ was about 20\% higher than $N_{\mathrm{EM}}$ at low photon levels. 
Because the power meter method provides a more direct measurement, we use it as a reference, adjusting the EM-CCD-derived photon counts as $1.2 N_{\mathrm{EM}}$.

\bigskip

\begin{backmatter}
\bmsection{Funding}
This work was supported by the Air Force Office of Scientific Research (AFOSR, grant FA9550-24-1-0345) and the Technology Innovation Program (IRIS no. RS-2024-00419426) funded by the Ministry of Trade, Industry and Energy (MOTIE, Korea).

\bmsection{Disclosures}
The authors declare no conflicts of interest.

\bmsection{Data Availability}
The datasets used in our paper, codes used for image reconstruction are available from the corresponding author upon request.

\bmsection{Supplemental document}
See Supplement 1 for supporting content.

\end{backmatter}


\bibliography{reference_correct}
\clearpage 
\appendix
\includepdf[pages=-]{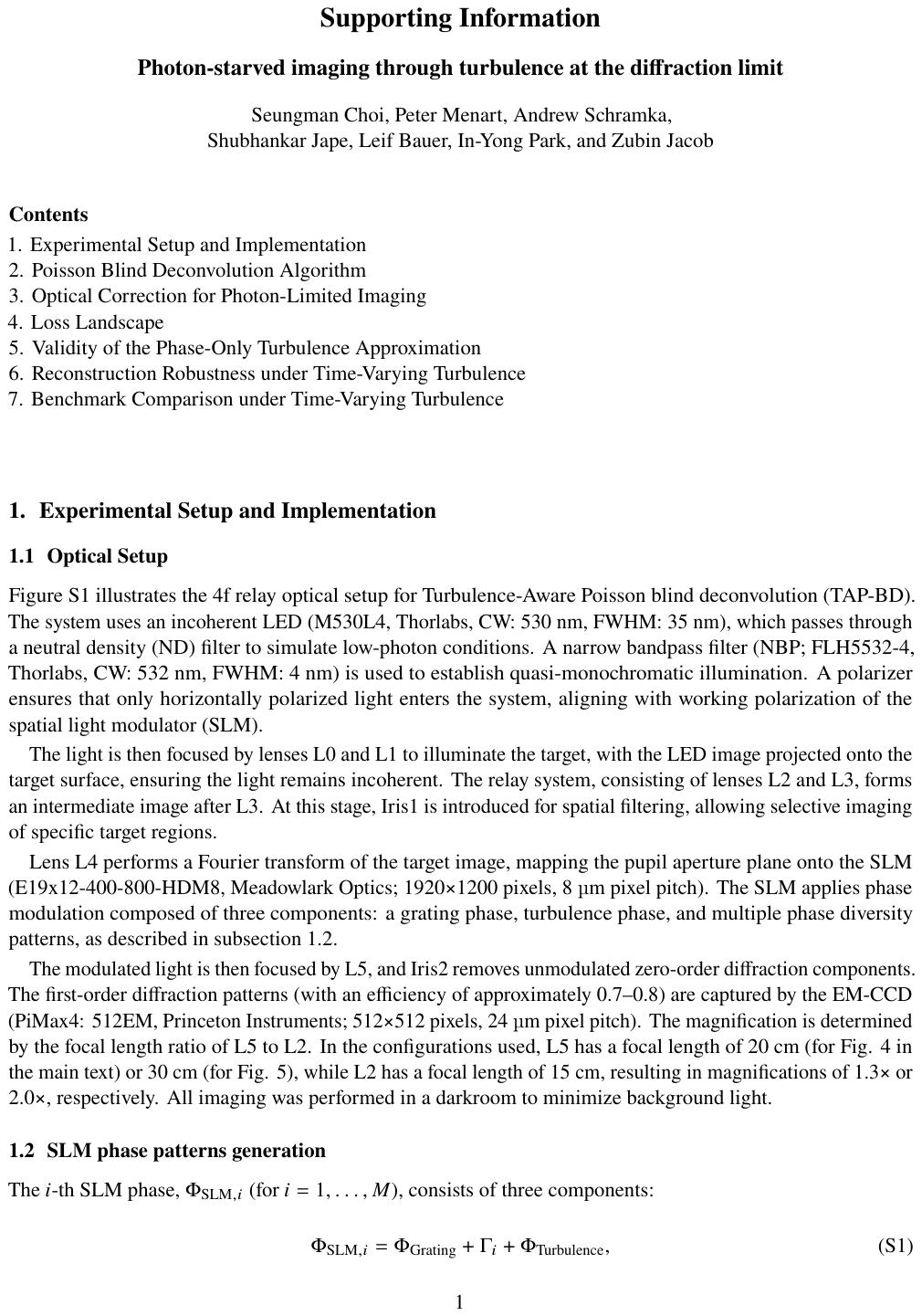}
\end{document}